\documentclass{emulateapj}
\usepackage{apjfonts}
\newcommand{\As}{\emph{Sat2Cen}}
\newcommand{\Bs}{\emph{KeepSat}}
\newcommand{\Cs}{\emph{Sat2ICL}}
\newcommand{\Ds}{\emph{Sat2Cen+ICL}}

\newcommand{\A}{\emph{Sat2Cen} }
\newcommand{\B}{\emph{KeepSat} }
\newcommand{\C}{\emph{Sat2ICL} }
\newcommand{\D}{\emph{Sat2Cen+ICL} } 

\begin{document}

\title{The Hierarchical Build-Up of Massive Galaxies and the
  Intracluster Light since $\lowercase{z}=1$}

\author{
Charlie Conroy\altaffilmark{1}, 
Risa H. Wechsler\altaffilmark{2},
Andrey V. Kravtsov\altaffilmark{3}
}

\slugcomment{Submitted to ApJ, 14 July 1789}

\altaffiltext{1}{Department of Astrophysical Sciences, Peyton Hall,
Princeton University, Princeton, NJ 08544}
\altaffiltext{2}{
  Kavli Institute for Particle Astrophysics and Cosmology, Physics Department,
and Stanford Linear Accelerator Center, 
  Stanford University, Stanford, CA 94305} 
\altaffiltext{3}{Department of Astronomy and Astrophysics,
  Kavli Institute for Cosmological Physics, \& 
  The Enrico Fermi Institute,
  The University of Chicago, 5640 S. Ellis Ave., Chicago, IL 60637}

\begin{abstract}

  We use a set of simulation-based models for the dissipationless
  evolution of galaxies since $z=1$ to constrain the fate of accreted
  satellites embedded in dark matter subhalos.  These models assign
  stellar mass to dark matter halos at $z=1$ by relating the observed
  galaxy stellar mass function (GSMF) to the halo+subhalo mass
  function monotonically.  The evolution of the stellar mass content
  is then followed using halo merger trees extracted from $N$-body
  simulations.  Our models are differentiated only in the fate
  assigned to satellite galaxies once subhalos, within which
  satellites are embedded, disrupt.  These models are confronted with
  the observed evolution in the massive end of the GSMF, the $z\sim0$
  brightest cluster galaxy (BCG)-cluster mass relation, and the
  combined BCG and intracluster light (ICL) luminosity distribution
  --- all observables expected to evolve approximately
  dissipationlessly since $z=1$. The combined observational
  constraints favor a model in which the vast majority ($\gtrsim80$\%)
  of satellite stars from disrupted subhalos go into the ICL
  (operationally defined here as light below a surface brightness cut
  of $\mu_i\approx 23$ mag arcsec$^{-2}$).  Conversely, models that
  leave behind a significant population of satellite galaxies once the
  subhalo has disrupted are strongly disfavored, as are models that
  put a significant fraction of satellite stars into the BCG. Our
  results show that observations of the ICL provide useful and unique
  constraints on models of galaxy merging and the dissipationless
  evolution of galaxies in groups and clusters.

\end{abstract}


\section{Introduction}

The formation and evolution of massive ($M_{\rm{star}}\gtrsim
10^{11} M_\odot$) elliptical galaxies is thought to be inexorably
linked to the formation and evolution of the large scale structure of
the Universe.  The classical picture wherein massive elliptical
galaxies form ``monolithically'' at $z>5$ \citep{Partridge67} has been
replaced by more nuanced scenarios that decouple the epoch at which
these galaxies formed the bulk of their stars from the epoch (or
epochs) at which these stars were assembled to form the final galaxy.
These more complex scenarios arise fairly naturally within the context
of the hierarchical growth of structure in the now favored
$\Lambda$CDM cosmology \citep[see e.g.][]{Baugh96, Neistein06}.  While
stellar population modeling has firmly placed the epoch of star
formation in these galaxies at $z>2$ \citep[e.g.][]{Bower92, Trager00,
  vanDokkum01, Thomas05, Jimenez06}, the assembly history of massive
galaxies is still far from clear, and is the focus of this work.

The evolving space density of massive galaxies over time provides
important clues to their assembly history.  Evolution in the galaxy
stellar mass function (GSMF) since $z\sim1$ appears quite mild for the
most massive galaxies \citep{Fontana04, Drory04, Bundy05, Borch06,
  Fontana06, Cimatti06,Andreon06}.  Estimates of evolution in the
luminosity function are also consistent with massive galaxies
passively evolving from $z\approx 1$ to the present
\citep{Cirasuolo06, Wake06, Faber06, Willmer06, Brown07, Caputi06,
  Blanton06}.  These same observations also find roughly a doubling in
the \emph{total} stellar mass density from $z\sim1$ to $z\sim0$ ---
the implication being that star formation occurs primarily in less
massive galaxies at $z<1$.  In addition to number counts, estimates of
the merger rate of massive galaxies can in principle constrain their
assembly history, though current observations fail to present a
consistent picture \citep{vanDokkum05,Bell06,Masjedi06}.

The majority of massive elliptical galaxies are (or have been in their
recent past) the brightest cluster galaxies within large group- or
cluster-sized halos (referred to simply as ``clusters'' in the
remainder of the paper), located near the centers of the halo
potential well. It is thus interesting to study the formation of such
galaxies in the general context of cluster formation.  If, as recent
observations suggest, the majority of massive galaxies were already in
place at $z\sim1$, then on the surface it appears difficult to
reconcile this with the much more substantial evolution of their host
dark matter halos (massive halos grow by factors of $\gtrsim3$ in mass
since $z=1$).  It is one of the goals of the present work to address
and resolve this tension.

Within the $\Lambda$CDM framework, groups and clusters of galaxies are
expected to be continually accreting new galaxies.  After entering a
cluster, the stars in satellite galaxies can 1) be all deposited onto
the central galaxy, 2) stay bound as a satellite galaxy, or 3) be
scattered into the intra-cluster light (ICL). For the purposes of this
study we define the ICL as the stars beyond the optical radius of the
central galaxy, i.e. the light not accounted for in the photometry of
the central galaxy itself.  In reality, a combination of these
possibilities can occur. For example, a satellite galaxy can be
partially stripped and deposit a fraction of its stellar mass into the
ICL before merging and contributing the rest of the stars to the
central galaxy. In addition, when two massive galaxies merge, a
certain fraction of stars will acquire large kinetic energy and will
move to large radii, outside the optical radius of the remnant. While
evolution in the space density of massive galaxies strongly constrains
the importance of the first scenario, this observation cannot readily
distinguish between scenarios two and three.

Fortunately, fates 2) and 3) have effects on observable properties of
the ICL, which can thus constrain the amount of stars that can be lost
to the ICL during mergers or tidal stripping.  These observations
suggest that the fraction of total cluster light bound up in the ICL
is $\sim10-30$\% \citep{Zibetti05, Gonzalez05, Krick06}, with the ICL
comprising a significant fraction ($\sim50-80$\%) of the combined
light from the central galaxy and ICL \citep[][note that this fraction
depends sensitively on the way that the ICL and BCG are
separated]{Gonzalez05, Seigar06}.  Additionally, models in which the
stellar component of satellite halos is never disrupted by tides would
have quite different predictions for the number of satellite galaxies
in a halo of a given mass, and thus for the small-scale clustering of
galaxies \citep[see e.g.][]{Berlind02}.  Indeed, the clustering of
massive galaxies in combination with their evolving space density has
recently been exploited by \citet{MWhite07} to constrain the
disruption rate of massive satellite galaxies between $z\simeq0.9$ and
$z\simeq0.5$.

The goal of this study is to confront these and other observational
constraints \emph{simultaneously} with a series of simple,
simulation-based models in order to gain insight into the fate(s) of
satellite stars.  The models presented herein combine a simple
prescription for relating galaxies to dark matter halos at $z\sim1$
with the assembly history of these halos extracted from $N$-body
simulations in order to follow the dissipationless growth of massive
galaxies to $z\sim0$.  The relation between galaxies and halos at
$z\sim1$ is generated by assigning the most massive galaxies to the
most massive halos monotonically. This has been shown to successfully
reproduce a wide variety of observations \citep{Colin99, Kravtsov99b,
  Neyrinck04, Kravtsov04, Vale04, Vale06, Tasitsiomi04, Conroy06a,
  Shankar06, Vale07}.  Note that this model considers both subhalos,
which are halos contained within the virial radii larger halos, and
what we will call distinct halos, which are halos not contained within
the virial radii of larger halos. We follow the dynamical evolution of
subhalos after they accrete onto their host halo using merger trees
extracted directly from cosmological simulations, rather than a
semi-analytic model.

Semi-analytic models (SAMs) for the formation and evolution of
galaxies within a cosmological context, depending on the adopted
assumptions about galaxy formation physics, are capable of predicting
both strong \citep{Baugh96, DeLucia06} and mild \citep{Bower06,
Kitzbichler07, Monaco06} evolution in the number density of massive
galaxies since $z\sim1$.  The most massive galaxies in many of these
models have formed the bulk of their stars at $z>2$, in agreement with
observations.  Hence differences between these models are due
primarily to different treatments of the assembly history of the
massive galaxies.

Our approach is similar in spirit to that of a recent study by
\citet{Monaco06} who used a SAM to follow the evolution of galaxies.
These authors artificially turned off star formation at $z<1$ in order
to follow the dissipationless growth of galaxies at late times,
similar to what we do here.  The orbital evolution of satellites in
their SAM was computed with simple analytical approximations to
dynamical friction, tidal heading and tidal stripping.
\citet{Monaco06} showed that the observed evolution in the space
density of massive galaxies is reproduced in their model only if they
allow for $>30$\% of stars from disrupted satellites to be transfered
into the ICL.  The present work goes further than the study of
\citet{Monaco06} by 1) using an independent set of simulations with
satellite tracks extracted directly from the simulations and 2)
comparing to a wider array of observations and hence providing more
general constraints on the fates of the stars within satellite
galaxies.

The rest of this article unfolds as follows.  In $\S$\ref{s:sims} we
describe the simulations, halos catalogs and merger trees used in this
analysis.  $\S$\ref{s:models} outlines the details of our models and
$\S$\ref{s:res} contains comparisons between the models and several
observations.  The implications of these results and comparison to
related work is discussed in $\S$\ref{s:disc}.  Throughout this paper
we assume a $\Lambda$CDM cosmology with $(\Omega_m, \Omega_{\Lambda},
h,\sigma_8) = (0.3,0.7,0.7,0.9)$, except in $\S$\ref{s:sims} where we
leave quantities in terms of the reduced Hubble constant, $h$.

\section{Simulations, Halo Catalogs, and Merger Trees}\label{s:sims}

The simulations used here were run with the Adaptive Refinement Tree
(ART) $N$-body code \citep{Kravtsov97,Kravtsov99a}.  The ART code
implements successive refinements in both the spatial grid and
temporal step in high density environments.  These simulations were
run in the concordance flat $\Lambda$CDM cosmology with
$\Omega_m=0.3=1-\Omega_{\Lambda}$, $h=0.7$, where $\Omega_m$ and
$\Omega_{\Lambda}$ are the present-day matter and vacuum densities in
units of the critical density, and $h$ is the Hubble parameter in
units of $100$ km s$^{-1}$ Mpc$^{-1}$.  The power spectra used to
generate the initial conditions for the simulations were determined
from a direct Boltzmann code calculation (courtesy of Wayne Hu).  We
use a power spectrum normalization of $\sigma_8=0.90$, where
$\sigma_8$ is the rms fluctuation in spheres of $8h^{-1}$ Mpc comoving
radius.

The simulation used herein was run in a box of length $120$ $h^{-1}$
Mpc with particle mass $m_p=1.07\times10^9$ $h^{-1}$ $M_\Sun$, peak
force resolution of $h_{peak}=1.8$ $h^{-1}$ kpc, and $512^3$
particles.  We have checked that our results remain unchanged when
utilizing a smaller box with smaller particle mass (a box of length $80$
$h^{-1}$ Mpc with particle mass $m_p=3.16\times10^8$ $h^{-1}$
$M_\Sun$).

From this simulation we generate dark matter halo catalogs
and dark matter halo merger trees.  Our models rely not only on
distinct halos, i.e., halos with centers that do not lie within any
larger virialized system, but also subhalos, which are located with
the virial radii of larger systems.  When we refer to a ``halo''
generically we mean both distinct halos and subhalos.


Distinct halos and subhalos are identified using a variant of the
bound density maxima (BDM) halo finding algorithm \citet{Klypin99}.
Details of the algorithm and parameters used can be found in
\citet{Kravtsov04}; we briefly summarize the main steps here.  All
particles are assigned a density using the \texttt{smooth}
algorithm\footnote{To calculate the density we use the publicly
  available code \texttt{smooth:
    http://www-hpcc.astro.washington.edu/tools/ tools.html}} which
uses a symmetric SPH smoothing kernel on the $32$ nearest neighbors.
Starting with the highest overdensity particle, we surround each
potential center by a sphere of radius $r_{\rm{find}}=50h^{-1}$ kpc
and exclude all particles within this sphere from further search.
Hence no two halos can be separated by less than $r_{\rm{find}}$.  We
then construct density, circular velocity, and velocity dispersion
profiles around each center, iteratively removing unbound particles as
described in \citet{Klypin99}.  Once unbound particles have been
removed, we measure quantities such as $V_{\rm{max}} =
\sqrt{GM(<r)/r}|_{\rm{max}}$, the maximum circular velocity of the
halo.  For each distinct halo we calculate the virial radius, defined
as the radius enclosing overdensity of $334$ with respect to the {\it
  mean} density of the Universe at the epoch of the output.  We use
this virial radius to classify objects into distinct halos and
subhalos.  The halo catalogs are complete for halos with more than
$50$ particles, which corresponds, for the box with length $120$
$h^{-1}$ Mpc, to $5.35\times 10^{10} h^{-1} M_\odot$.

Halo merger trees have also been constructed for this simulation
\citep[for a detailed description of the merger tree construction,
see][]{Allgood05}.  These merger trees allow us to tabulate
$M_{\rm{vir}}^{\rm{acc}}$ for subhalos, the virial mass at the time
when a subhalo first crosses the virial radius of a distinct halo.
Since subhalos are subject to dynamical processes such as tidal
stripping, $M_{\rm{vir}}^{\rm{acc}}$ will always be greater than or
equal to the present $M_{\rm{vir}}$.  This accretion epoch quantity is
used in our models, to which we now turn.

\section{The Models}\label{s:models}

\subsection{Connecting galaxies to halos}

Recent studies have shown that models in which galaxies are associated
with the centers of dark matter halos and subhalos accurately
reproduce a wide variety of observations both at low and high redshift
\citep[e.g.][]{Kravtsov04, Vale04, Conroy06a}.  Herein, stellar mass
is assigned to each halo in a simulation by assuming a monotonic
relation between stellar mass and halo virial mass using the observed
galaxy stellar mass function (GSMF) and halo mass function measured in
simulations:
\begin{equation}\label{eqn:v2l}
  n_g(>{M_{\rm{star},i}}) = n_h(>{M_{\rm{vir},i}})
\end{equation}
where $n_g$ and $n_h$ are the number density of galaxies and halos
(note again that ``halos'' here and throughout refers to both distinct
halos and their subhalos), respectively.  Galaxy stellar masses can
hence be assigned to halos at any epoch once the GSMF at that epoch is
known.  In the simplest version of this scheme there is assumed to be
no scatter in the relation between halo mass and stellar mass (see
$\S$\ref{s:robust} for a discussion of scatter in the context of our
models).

Subhalos lose mass due to tidal stripping as they orbit within their
parent halo.  Since stripping primarily affects the outer regions of
the subhalo, we expect the galaxy, which resides within the inner few
kpc of the subhalo, to be relatively unaffected by this
process. Hence, for subhalos, when relating halo mass to luminosity or
stellar mass we use its virial mass at the epoch when it is first
accreted onto the parent halo, $M_{\rm{vir}}^{\rm{acc}}$, rather than
its mass at the epoch of observation.  This choice is well motivated
both by hydrodynamical simulations \citep{Nagai05} and detailed
modeling of the small to intermediate scale ($0.1<r<10$ $h^{-1}$ Mpc)
clustering of galaxies over a range of redshifts \citep{Conroy06a}.

One may ask to what extent it is justifiable to identify satellite
galaxies with subhalos in dissipationless simulations.  It has been
shown that the subhalo population in dissipationless simulations is
indeed quite similar to the galaxy population in hydrodynamical
simulations \citep{Zheng05, Weinberg06} and semi-analytic models
\citep{Zheng05}. In particular, satellite populations in these
hydrodynamical simulations with cooling and galaxy formation have an
almost identical halo occupation distribution to the subhalos in
dissipationless simulations.  These conclusions are corroborated by
the general success of the subhalo-based models of galaxy clustering.

This model, for example, accurately captures observed relations
between cluster luminosity and the number of galaxies within a cluster
as a function of cluster mass \citep{Vale04, Vale06}, the luminosity
dependence of the galaxy-matter cross-correlation function
\citep[after a reasonable amount of scatter is introduced into this
relation; see][for details]{Tasitsiomi04}, close pair counts
\citep{Berrier06}, the luminosity, scale, and redshift-dependence
of the galaxy autocorrelation function from $z\sim5$ to $z\sim0$
\citep{Conroy06a}, and mass-to-light ratios in local clusters 
\citep{Tasitsiomi07}.

Emboldened by the success of this simple model, in the present work we
extend it by populating halos with galaxies at $z\sim1$ in the above
way and then using halo merger trees derived from $N$-body simulations
to follow the evolution of these galaxies to $z\sim0$.  Specifically,
we use the observed galaxy stellar mass function (GSMF) of
\citet{Fontana06} at $z\sim1$ to assign stellar masses to halos, and
then follow the \emph{dissipationless} evolution of these galaxies via
the merging history of their dark matter halos to $z\sim0$.  This
exercise is appropriate for the evolution of the most massive
galaxies, as these galaxies formed the bulk of their stars at $z>2$
\citep[e.g.][]{Bower92, Trager00, vanDokkum01, Thomas05, Jimenez06}
and hence largely evolve dissipationlessly at $z<1$. Note that
neglecting star formation at $z<1$ means that the evolution of the
total stellar mass predicted by our models is a lower limit on the
actual evolution.

The following $z=1$ GSMF Schechter parameters are adopted from
\citet{Fontana06}:
\begin{equation}
\begin{array}{ccl}
\alpha &= &-1.26 \pm0.1 \\
M_\ast& =& 10^{11.01\pm0.1} \,\,\, M_\odot\\
\phi_\ast& = &(7.6\pm2.4) \times 10^{-4} \,\,\, \rm{Mpc}^{-3}. 
\label{e:sparams}
\end{array}
\end{equation}
The $1\sigma$ errors quoted above are not the statistical errors
reported in \citet{Fontana06} but are instead meant to roughly
encompass the various published estimates of the $z=1$ GSMF Schechter
parameters.  The statistical errors are a factor of $2-3$ smaller than
these approximate systematic errors.  Here and throughout the Chabrier
IMF is used when quoting stellar masses.

It is important to note that while different authors derive somewhat
different Schechter parameters for the $z\sim1$ GSMF, all authors
agree that the massive end of the GSMF ($>M_\ast$) evolves very
little, if at all, since $z\sim1$.  In addition, while several
measurements of the GSMF at $z\sim1$ have relied on photometric
redshifts \citep[e.g.,][]{Borch06}, the general conclusions from these
studies have been supported by measurements which utilize
spectroscopic redshifts \citep{Bundy06,Fontana06}.  

\begin{figure}
\plotone{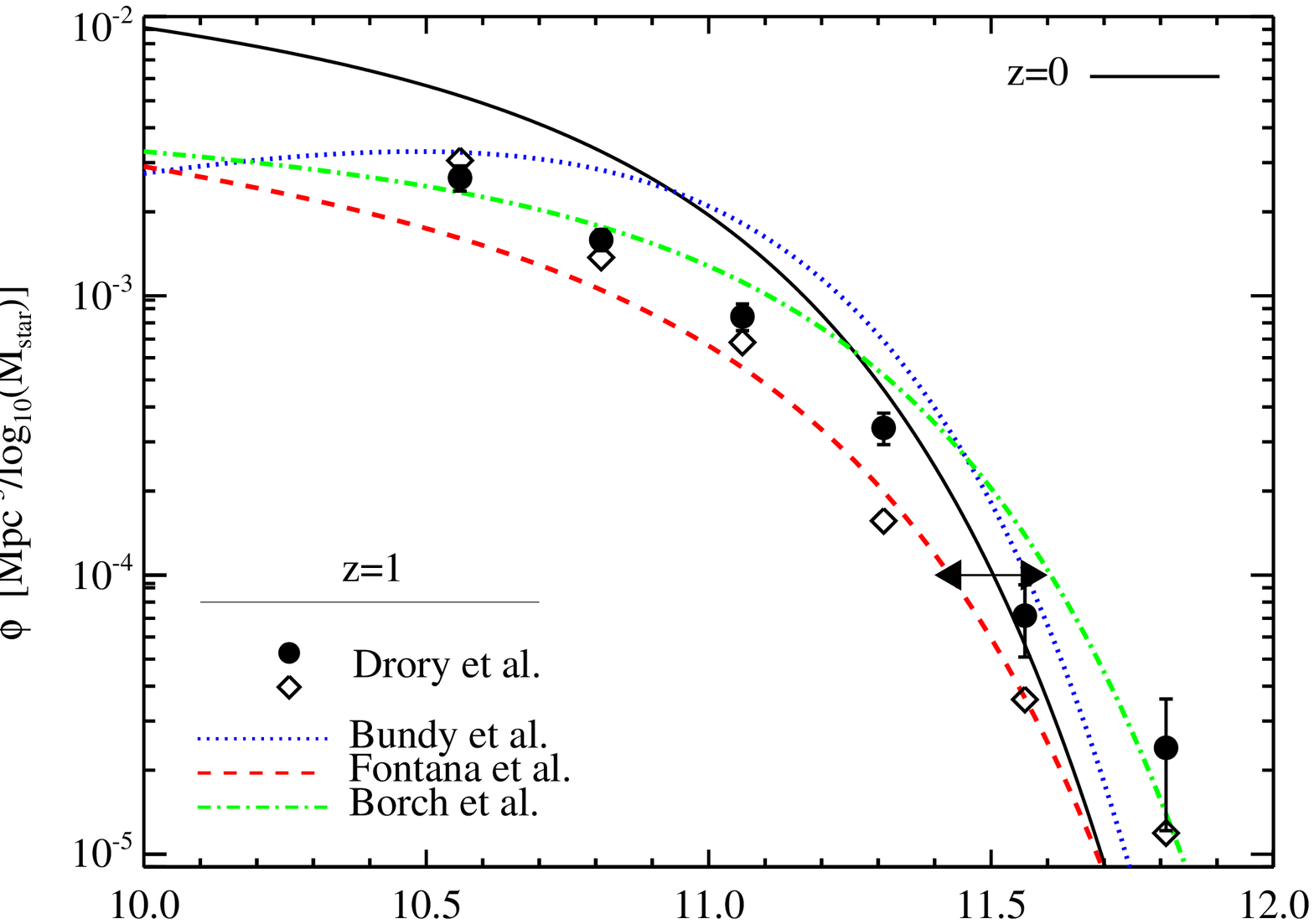}
\vspace{0.5cm}
\caption{Observed galaxy stellar mass functions (GSMFs).  The
  non-solid lines are best fit Schechter functions to the following
  GSMFs at $z\sim1$: \citet[][ \emph{dot-dashed line}]{Borch06},
  \citet[][ \emph{dotted line}]{Bundy06}, \citet[][ \emph{dashed
    line}]{Fontana06}.  The GSMF from \citet{Drory04} is plotted
  directly, both for their fiducial GSMF (\emph{diamonds}) and an
  estimate which includes lost-light corrections (\emph{solid
    circles}; error bars are Poisson uncertainties only).  The GSMF at
  $z\sim0$ from \citet[][ \emph{solid line}]{Cole01} is also included,
  along with arrows indicating a change of 0.2dex in $M_{\rm{star}}$
  at $\phi=10^{-4}$.}
\vspace{0.5cm}
\label{f:obs}
\end{figure}

Figure \ref{f:obs} presents a comparison between various observed
GSMFs at $z\sim1$ and the $z\sim0$ GSMF from \citet{Cole01}.  Note
that the $z\sim1$ GSMF used in our models \citep{Fontana06} is below
all the other $z\sim1$ GSMFs.  This implies that our results
concerning the $z=0$ brightest cluster galaxy (BCG) luminosities
derived with this GSMF will be lower bounds relative to the other
GSMFs.

At $z=1$, a space density of $10^{-4}$ Mpc$^{-3}$ corresponds to halos
of virial mass $M_{\rm{vir}}\sim 4\times 10^{13} M_\odot$ and stellar mass
$M_{\rm{star}}\sim 3\times 10^{11} M_\odot$.  At this space density,
$\sim17$\% of halos are subhalos at $z\sim0$.  Finally, note that
although we formally assign stellar mass to all the halos found in the
simulations and track their evolution as described in the following
section, our results are quite insensitive to low mass halos and are
instead governed by the evolution and fate of much more massive halos.
Since it is these halos that are most easily resolved and tracked from
timestep to timestep, we expect our results to be insensitive to our
simulation resolution.

\subsection{Dynamical evolution models}

Our model contains (at least) three adjustable components.  The first
is our assumed cosmology and will be fixed throughout, though we
comment qualitatively on the effects of changing certain cosmological
parameters in $\S$\ref{s:robust}.  The second component is our input
method for assigning stellar masses to halos at $z\sim1$.  In
$\S$\ref{s:robust} we demonstrate that our results are robust to
reasonable changes of this second component.  In particular, we
introduce scatter in the assignment between stellar mass and halo mass
and marginalize over the uncertainties in the $z\sim1$ GSMF and find
no qualitative change to our conclusions, provided that the true GSMF
is not considerably different from recent estimates.

The third adjustable component is the most uncertain and concerns how
we treat subhalos that have dropped out of the halo catalog.  This can
occur either because the subhalo is physically disrupted or because it
is stripped below the resolution limit of the simulation; it is often
quite difficult to distinguish these two cases within the simulation.
We do nothing to the galaxies assigned to subhalos while the subhalos
remain identifiable in our simulation --- i.e., the satellite galaxy
within the subhalo experiences no tidal stripping.  Once a subhalo is
destroyed, we are free to redistribute the stars from this subhalo in
one or more of the ways outlined in the Introduction.  To summarize
that discussion: the stars can be deposited onto the central galaxy
(this assumes that the satellite within the destroyed subhalo has
merged with the central galaxy), can remain as a satellite galaxy
without an identifiable subhalo, and/or the stars can be added to the
intra-cluster light (ICL).  In the latter case the stars are added to
the outer regions of the central galaxy, beyond the optical radius
(the radius within which the central galaxy luminosity is measured).

In order to explore these possibilities, four models are constructed
which differ only in the fate of the stars within disrupted subhalos.
Model \A assumes that all of the stars are deposited onto the central
galaxy.  Conversely, model \B assumes that the stars remain bound as a
satellite galaxy.  Model \C assumes that all stars are deposited into
the ICL.  Finally, model \D assumes that the stars are distributed
equally between the ICL and the central galaxy.  We will assume below,
when comparing to observations of the ICL, that the ICL in these
models is generated predominantly from the remnants of mergers with the
central galaxy at $z<1$.  These assumptions are motivated by
hydrodynamical simulations of clusters \citep{Willman04, Murante04,
  Rudick06, Murante07, SommerLarsen05} which showed that the majority
of the ICL is built up at $z<1$ from major mergers with the central
galaxy, rather than tidal stripping as the satellite orbits in the
cluster.  For reference, these models are summarized in Table
\ref{t:models}.

These models have implicitly assumed that all stellar mass at $z=1$
persists to $z=0$.  However, stellar mass-loss due to winds and
supernovae can result in a significant decrease in the aggregate
stellar mass of a population over time.  In order to understand these
effects, consider a secular mass-loss rate of $\dot{M}/M=0.05\,
(t/\rm{Gyr})^{-1}$ for a stellar population formed in an instantaneous
burst and older than a few hundred Myr \citep{Jungwiert01}.  If
massive red galaxies grew most of their stellar mass in the form of a
single burst at $z=2$, then the fraction of stellar mass lost between
$z=1$ and $z=0$ is only $7$\%; however, if the stars in these galaxies
all formed at $z=1$ then the fraction is $36$\%.  Since observations
place the epoch of star-formation in these massive galaxies at
$z\gtrsim2$, mass-loss effects are likely unimportant.

\begin{deluxetable}{ll}
\tablecaption{Summary of Models}
\tablehead{ \colhead{Model} & \colhead{Fate of satellite galaxy}\\
\colhead{} & \colhead{in a disrupted subhalo} }
\startdata\\
\A & Stars deposited onto the central galaxy \\
\B & Stars remain bound as a satellite galaxy \\
\C & Stars deposited into the ICL \\
\D & Stars divided equally between the ICL \\
  &   and the central galaxy \\
\enddata
\label{t:models}
\end{deluxetable}

\subsection{Generating luminosities}

Comparison to observations in $\S$\ref{s:BCG} and $\S$\ref{s:icl} will
require conversion from stellar masses to $K$-band and $I$-band
luminosities, respectively.  To make this conversion we use the
relation between mass-to-light ratios and colors provided by
\citet{Bell03},
\begin{equation}
\begin{array}{ccc}
M_{\rm{star}}/L_K &=& 0.72 \\
M_{\rm{star}}/L_I &=& 1.90 \\
\end{array}
\label{e:ml}
\end{equation}
where we have assumed a color of $(u-r)=2.5$ and $(B-R)=1.5$ when
deriving $M_{\rm{star}}/L_K$ and $M_{\rm{star}}/L_I$, respectively,
and a Chabrier IMF. These colors are appropriate for the bright end of
the red sequence \citep{Pahre99,Baldry04, Bell03}.  

All the results to be discussed below are independent of our assumed
IMF because we are only interested in relative evolution from $z=1$ to
$z=0$. Thus, so long as the observations at these epochs use the
same IMF, the results are insensitive to the particular IMF used
(whether for example Chabrier, Kroupa, or Salpeter IMFs are used).
Results which concern luminosities are also IMF independent so long as
we use a mass-to-light ratio with the same IMF as used in our GSMFs
(as we have done above).

The luminosities of galaxies at $z\sim0$ in these models are strictly
lower bounds for two reasons: 1) these models neglect star-formation
since $z=1$, and 2) some galaxies might not be as red as the colors
assumed in the previous paragraph, and hence they will be brighter at
a fixed stellar mass.  For the massive BCGs studied herein however,
these effects are unimportant.  It will become apparent in the next
section that including these possibilities would only strengthen our
general conclusions since both residual star-formation and bluer
colors would increase the BCG luminosities and the evolution of the
massive end of the GSMF.

We also assume that the ICL has the same color as the tip of the red
sequence when converting ICL stellar mass to luminosities.  Such an
assumption appears borne out by observations of an at most weak color
gradient out to several hundred kpc from the BCG, with some authors
finding a slight reddening \citep{Gonzalez00, Krick06} and others a
slightly bluer color, or no gradient at all \citep{Zibetti05}.

\section{Results}\label{s:res}

We now compare the models constructed in $\S$\ref{s:models} to the
observed evolution in the galaxy stellar mass function
($\S$\ref{s:evgsmf}), to the relation between BCG luminosity and
cluster virial mass at $z\sim0$ ($\S$\ref{s:BCG}), and to properties
of the intracluster light (ICL; $\S$\ref{s:icl}).  In $\S$\ref{s:star}
we discuss these models in the context of star formation since $z=1$;
at the end of this section several caveats and assumptions made herein
are explored.

\subsection{Evolution in the GSMF}\label{s:evgsmf}

\begin{figure}
\plotone{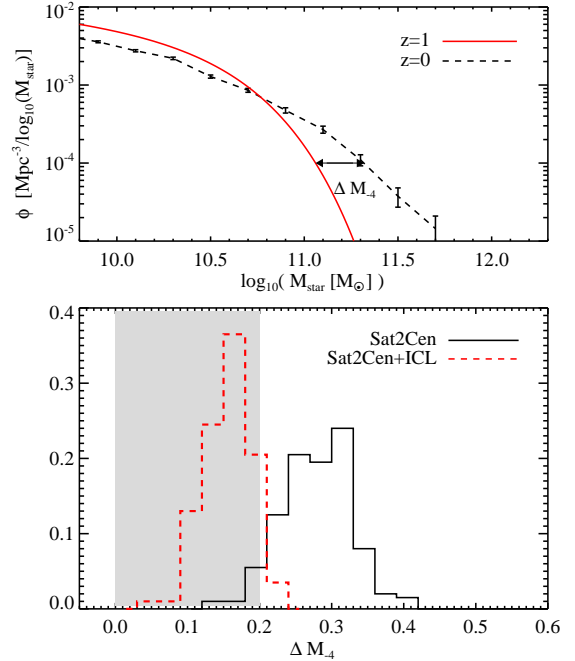}
\vspace{0.5cm}
\caption{Evolution in the GSMF from $z\sim1$ to $z\sim0$.  \emph{Top
    panel}: Pictorial representation of how we quantify evolution in
  the massive end of the GSMF with the parameter $\Delta M_{-4}$,
  for one realization of model \A (error bars denote Poisson
  uncertainty only).  See the text for details.  \emph{Bottom panel}:
  Distributions of $\Delta M_{-4}$ for models \A and \D.  These
  distributions are obtained by running each model using a slightly
  different GSMF to populate halos at $z\sim1$.  Model \B predicts no
  change in the entire GSMF and hence $\Delta M_{-4}\equiv 0$ for
  that model, while model \C results in $\Delta M_{-4}\sim0$.
  Running models for a series of GSMFs demonstrates the effect of
  uncertainties in the $z\sim1$ GSMF on our results. 
  Observational limits are denoted by the shaded region.}
\vspace{0.5cm}
\label{f:m45}
\end{figure}

In this section we present the evolution in the galaxy stellar mass
function (GSMF) for the models described in $\S$\ref{s:models}, and
compare to observations.  An effective way to quantify the evolution
in the massive end of the GSMF is by quantifying the change of stellar
mass corresponding to galaxies with a fixed given value of spatial
number density.  Hence, we define the quantity $M_{-4}$, such that
$\phi(M_{-4})=10^{-4}$, and its evolution as $\Delta M_{-4}\equiv
M_{-4}^{z=0}- M_{-4}^{z=1}$ (see the top panel of Figure \ref{f:m45}
for a pictorial representation).  This quantity is more stable than
its inverse (the evolution in the number density of a given stellar
mass) due to the exponential cut-off of the GSMF at high stellar
masses. Note that $M_{-4}$ is dominated by the space density of
massive, rare objects.

Uncertainties in the observed $z\sim1$ GSMF are incorporated into the
models by generating $200$ realizations of the GSMF with each
Schechter parameter drawn from a Gaussian distribution with mean and
dispersion equal to the best fit and $1\sigma$ errors on the observed
GSMF (see Equation \ref{e:sparams}).  We have tried other GSMFs at
$z\sim1$ \citep{Bundy06, Borch06} and find qualitatively similar
results.

The bottom panel of Figure \ref{f:m45} shows the distribution of
$\Delta M_{-4}$ generated from the $200$ Monte-Carlo realizations for
models \A and \D (models \B and \C produce no/little change in $\Delta
M_{-4}$, see below).  The observational limits are denoted by the
shaded band and have been estimated from observations of the $z=1$
GSMF (see Figure \ref{f:obs}).  Note that for each model the predicted
evolution is a lower bound, since star-formation between $z\sim1$ and
$z\sim0$, which is neglected in these models, will increase $\Delta
M_{-4}$.  However, since the massive galaxies which dominate this
quantity have formed the bulk of their stars at $z>1$, we expect the
contribution to $\Delta M_{-4}$ from star-formation to be unimportant.
Note that the evolution in the GSMF presented in Figure \ref{f:m45}
makes no reference to the observed $z\sim0$ GSMF.  We simply compare
the observed GSMF at $z=1$ with the GSMF evolved to $z=0$ with our
models.  For comparison, $\Delta M_{-4}=0.4$ for dark matter halos in
our adopted cosmology.

For each model, the differences in $\Delta M_{-4}$ for different
$z\sim1$ GSMF Schechter parameters arise because different Schechter
parameters result in a different relation between stellar mass and
dark matter halo mass, via Equation \ref{eqn:v2l}.  The growth of a
dark matter halo and, through our models, the growth of the central
galaxy, is driven by accreted halos spanning a range in mass.
Therefore, the predicted growth of the central galaxy mass will
change, if the mapping between halo mass and stellar mass changes.  We
now explain the behavior of each model in turn.

Model \A displays the largest increase in $M_{-4}$ because massive
galaxies, which by definition dominate this quantity, are growing
rapidly.  In this model, galaxies with $M_{\rm{star}}> 10^{11} M_\Sun$
at $z=0$ have on average more than doubled in mass since $z=1$.  Rapid
growth of massive galaxies occurs in this case because the satellite
galaxies within disrupted subhalos add all of their stars onto the
central massive galaxy.  Model \B predicts $\Delta M_{-4}\equiv0$,
since in this model galaxies do not evolve, i.e., galaxies neither
merge nor form stars since $z=1$.  In Model \C the massive end of the
GSMF does not increase (but in some realizations actually decreases
slightly with time as some very massive satellite galaxies disrupt)
because the stars within satellite galaxies are transfered to the ICL,
which, for the purposes of the GSMF amounts to deleting the galaxy
from the sample. Model \D is, by construction, intermediate between
models \A and \Cs, since half of the stars from disrupted subhalos are
deposited onto the central galaxy and the rest into the ICL.  In this
model the most massive galaxies have increased in mass by the more
modest factor of $\sim50$\% since $z=1$.

Based on the evolution in the GSMF, model \A is strongly disfavored.
Models \Bs, \Cs, and \D fair far better.  In fact, based on current
observations, all of these models appear more or less equally viable
(their relative viability depends on what one assumes about the
observationally allowed range in $\Delta M_{-4}$).  We now turn to
comparisons of the models with observations of BCGs and the ICL, in
the hope of more strongly distinguishing between them.

\begin{figure}[t!]
\plotone{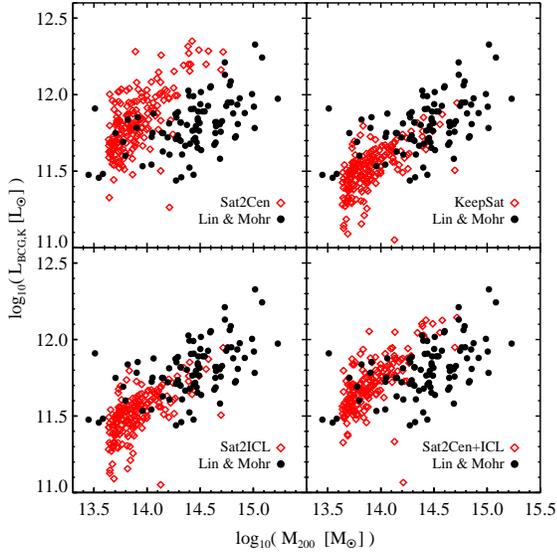}
\vspace{0.5cm}
\caption{Luminosity of the BCG as a function of the cluster virial
  mass, comparing data from \citet{Lin04b} (\emph{solid circles}) to
  models (\emph{open diamonds}).}
\vspace{0.5cm}
\label{f:bcg}
\end{figure}

\subsection{BCG Luminosities at $z\sim0$}\label{s:BCG}

We now confront our models with observations of cluster properties at
$z\sim0$.  \citet{Lin04b} have computed virial masses from $X$-ray
observations and, using the 2MASS database, estimated BCG
luminosities\footnote{Note that $\sim70$\% of the BCGs in this sample
  are centered in the cluster to within $5$\% of the virial radius; in
  what follows we assume that all BCGs in this sample are the central
  galaxy.} for $93$ clusters at $z<0.1$.  Figure \ref{f:bcg} plots the
$K$-band BCG luminosity versus cluster virial mass for the data from
\citet[][\emph{solid circles}]{Lin04b} and for the models\footnote{For
  this comparison, we have converted our halo mass definition of 334
  times the mean density of the Universe to the definition used in
  \citet[][200 times the critical density]{Lin04b}, by using an NFW
  density profile with a mass-dependent concentration.}.  Several
trends are apparent.  Models \B and \C predict identical BCG
luminosities (since in both models the central galaxy, which is
identified as the BCG, does not accrete any satellite stars since
$z=1$) and are in good agreement with the observations.  Model \A is,
again, in strong disagreement with the observations.  Finally, the
predictions of model \D are in between those of models \B and \C and
model \As, by construction, and are mildly disfavored by the
observations.

The failure of model \A is of course no surprise in light of the
results in $\S$\ref{s:evgsmf}.  The failure is, as in $\S$\ref{s:BCG},
simply a manifestation of the fact that the massive end of the halo
mass function in a $\Lambda$CDM cosmology evolves much more strongly
from $z=1$ to $z=0$ than the observed evolution in the GSMF.  This is
corroborated by observational constraints on halo masses at various
epochs which indicate that while the stellar and dark matter
components grow in lock-step for lower-mass systems
\citep{Heymans06,Conroy07}, the stellar mass growth of central
galaxies in high mass halos appears to be outpaced by the growth in
their halo mass at $z<1$ \citep{Conroy07}.

Comparison to BCG luminosities does however provide stronger
constraints on model \D compared to the constraints from evolution in
the GSMF.  Specifically, the disagreement between observations
apparent in Figure \ref{f:bcg} suggests that less than half of the
stars from disrupted subhalos can end up in the central BCG.  And
indeed, from the agreement between observations and models \B and \Cs,
we conclude that no growth is favored.

\begin{figure}
\plotone{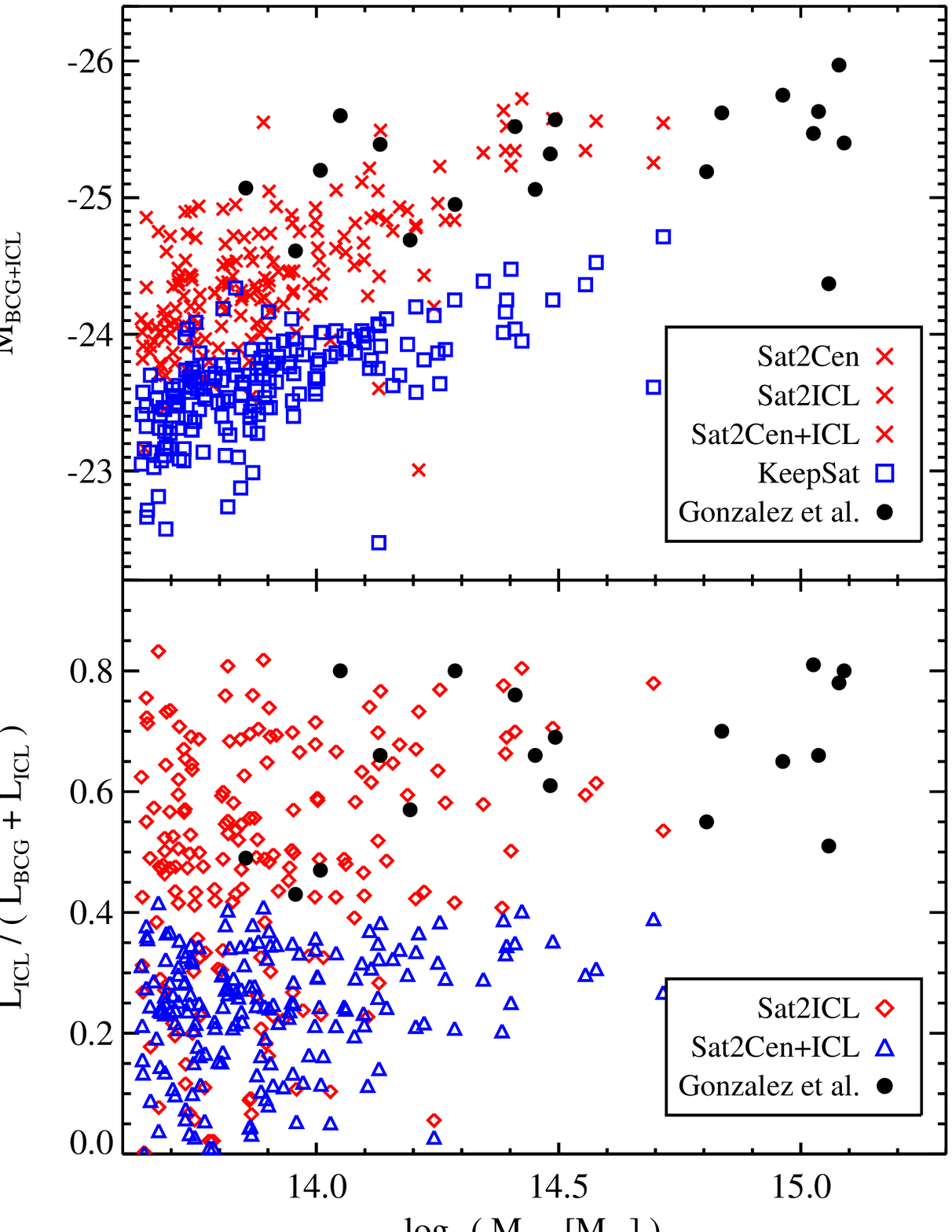}
\vspace{0.5cm}
\caption{\emph{Top panel:} Absolute $I$-band magnitudes of the
  combined BCG and ICL components as a function of total cluster
  mass.  Observations at $z\sim0$ \citep[][\emph{solid
    circles}]{Gonzalez05,Gonzalez07} are compared to models \B
  (\emph{squares}), \As, \Cs, and \D (\emph{crosses}).  \emph{Bottom
    panel:} fraction of BCG+ICL luminosity that is contained in the
  ICL component as a function of cluster virial mass, comparing
  observations to models \C (\emph{diamonds}) and \D
  (\emph{triangles}).  Models \A and \B produce no ICL component.}
\vspace{0.5cm}
\label{f:ficl}
\end{figure}

\subsection{The ICL Component at $z\sim0$}\label{s:icl}

We now confront our models with observations of the
ICL component.  The notion of intracluster light arose from the
observation that the extended profiles of BCGs were in excess of de
Vaucouleurs profiles \citep{Matthews64,Schombert88}.  There is
currently no strong consensus on whether the ICL is simply the outer
component of the BCG or whether it is dynamically distinct.
Observationally, the ICL is often defined as the total light beyond a
particular surface brightness level, although recently there have been
attempts to model the entire surface brightness profile with multiple
components, thus separating the BCG and ICL in a less arbitrary way.
For our purposes we use the data from \citet{Gonzalez05,Gonzalez07}
who have measured the surface brightness profiles for $24$ BCGs at
$z<0.12$ in the $I$-band and have also measured virial masses for the
clusters.

When considering the ICL, the most straightforward observable to
confront with models is the combined BCG and ICL light.  For this
quantity one does not need to rely on the potentially arbitrary
distinction between BCG and ICL.  The top panel of Figure \ref{f:ficl}
plots the absolute $I$-band magnitude for the combined BCG and ICL
components, $M_{\rm{BCG+ICL}}$, as a function of cluster virial
mass\footnote{Their definition of virial mass is the mass enclosing a
  region with mean density equal to 500 times the critical density; we
  have converted both their masses and our to a definition of 200 times
  the critical density; see $\S$\ref{s:BCG} for details.}, for the data
from \citet[][\emph{solid circles}]{Gonzalez05}, and for our models.
Models \As, \Cs, and \D predict the same $M_{\rm{BCG+ICL}}$, since
these models differ only in the way in which the stars are distributed
between the BCG and ICL.  As can be seen from the figure, these three
models all adequately reproduce the observations over a range of
cluster masses.

Model \B however predicts a substantially different
$M_{\rm{BCG+ICL}}$, since in this model no stars are added to the ICL
nor BCG since $z=1$.  In particular, model \B predicts
$M_{\rm{BCG+ICL}}$ $>1$mag lower than observations, which corresponds
to a factor $>2.5$ lower luminosity, and because of our adopted
constant mass-to-light ratio, this corresponds to the same factor
lower in stellar mass.  This discrepancy is far too great to be
accounted for by the small effects neglected in these models, such as
star-formation, tidal stripping, and ICL generation at $z>1$ (see
below).  Hence, observations of $M_{\rm{BCG+ICL}}$ strongly suggest
that model \B is unrealistic.

A more complex, but potentially more discriminating observable, is the
fraction of BCG and ICL light that is in the ICL.  In this case
comparison between models and data must be treated carefully because
the separation between ICL and BCG is not handled in the same way for
different datasets.  Our operational definition of ICL is simply the
light not counted as the BCG by \citet{Lin04b}.  Their definition of
BCG luminosity corresponds to the light within a surface brightness of
$\mu_K\approx 21$ mag arcsec$^{-2}$.  Assuming $I-K=2$, which is
appropriate for bright red galaxies, implies a separation between ICL
and BCG at $\mu_I= 23$ mag arcsec$^{-2}$.  The observational results
from \citet{Gonzalez05}, which are in the $I$-band, have been recast
in this way to afford the most robust comparison to our model
(A. Gonzalez, private communication).

The bottom panel of Figure \ref{f:ficl} plots the ICL fraction for
models \C and \D (recall that models \A and \B do not have an ICL
component) and compares to the results from \citet[][\emph{solid
  circles}]{Gonzalez05}.  It is clear that model \C predicts much more
ICL light than the other models and an ICL light fraction that is in
excellent agreement with the observations.

There are two additional routes by which the ICL can be built up that
have been neglected thus far: build up of the ICL at $z>1$ and the
tidal stripping of satellites as they orbit within the cluster
potential.  Specifically, hydrodynamical simulations have found that
$\sim85$\% of the stars in the ICL at $z=0$ were deposited at $z<1$
\citep{Willman04, Murante07}, and less than $30$\% of the ICL was
built up by tidal stripping of satellite galaxies \citep{Murante07}.
The majority of the ICL is thus built during violent merging events
with the central galaxy and/or the complete disruption of satellites
--- i.e. the two processes that are captured in our treatment of the
ICL. Recent observations of the color of the ICL support the picture
that stars comprising the ICL formed at $z>1$ \citep{Krick06}.  While
both processes will increase the ICL component, neither will produce
enough additional ICL to account for the discrepancy between models
\As, \B and \D and observations depicted in Figure \ref{f:ficl} --- if
these simulations are accurately capturing the build-up of the ICL.

\subsection{Implications for Star Formation Since $z=1$}\label{s:star}

Until now we have focused on observations that can be described with
purely dissipationless modeling.  Now that we have identified a
dissipationless model that adequately reproduces various observations
(model \Cs), we can ask what more must be added to such a model in
order to reproduce the observed global galaxy population.  Clearly,
the most important process neglected thus far is star formation, which
becomes increasingly important in lower mass halos.  We now turn to a
discussion of the importance of star formation as a function of halo
mass since $z=1$.

\begin{figure}
  \plotone{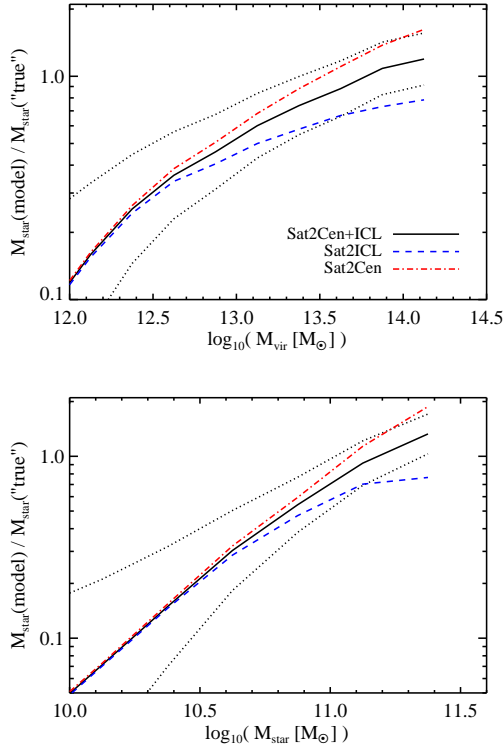}
\vspace{0.5cm}
\caption{Ratio of the $z=0$ stellar masses predicted by models \As, \C
  and \D to the ``true'' stellar masses, as a function of $z=0$ host
  halo mass (\emph{top panel}) and stellar mass (\emph{bottom panel}).
  These ``true'' stellar masses are obtained by matching the observed
  $z=0$ GSMF to the $z=0$ halo mass function (see the text for
  details).  Dotted lines denote the $1\sigma$ dispersion around model
  \D estimated from 200 Monte Carlo realizations that incorporate the
  $z=1$ GSMF uncertainties (the dispersion around the other models is
  similar).}
\vspace{0.5cm}
\label{f:star}
\end{figure}

We first construct a simple model that places the ``true'' $z=0$
stellar mass in dark matter halos.  This is accomplished using the
methodology outlined in $\S$\ref{s:models}, now matching the $z=0$
GSMF to the $z=0$ halo mass function.  Such a model will have the
correct $z=0$ GSMF by construction and should have approximately the
correct relation between stellar mass and halo mass since this method
has been shown to reproduce numerous observations remarkably well (see
$\S$\ref{s:models} for details).

Figure \ref{f:star} compares these true stellar masses to stellar
masses from models \As, \C and \D as a function of $z=0$ halo mass
(top panel) and stellar mass (bottom panel).  As before, we generate
200 realizations of these models by sampling the $z=1$ GSMF
uncertainties (recall that in these models the $z=0$ stellar masses
are products of the $z=1$ GSMF combined with the dark matter halo
merger trees to $z=0$).  The resulting mean and $1\sigma$ dispersions
are included in this figure.  At large masses, the stellar masses from
models \C and \D match the ``true'' stellar masses while model \A
overpredicts the true stellar masses, although all are consistent with
the true masses at roughly the $2\sigma$ level.  This is not
surprising, both in light of the results from previous sections and
more generally because massive galaxies (which reside in massive
halos) are observed to have finished forming stars by $z\approx2$.  In
fact, if there is truly zero star-formation in these massive galaxies
since $z=1$, then Figure \ref{f:star} suggests that the $z=0$ ``true''
stellar masses are best reproduced by a model in between models \C and
\Ds, i.e. subhalos transfer perhaps $\sim80$\% of their stars to the
ICL, and $\sim20$\% to the central galaxy.  A distinction this refined
should, of course, be treated with caution.

The behavior at lower masses is more interesting.  In this regime the
stellar masses from models \As, \C and \D are substantially less than
the true masses, indicating that the $z<1$ star formation is
increasingly important in halos of lower mass.  In fact, according to
these models roughly $40$\% of the stars in $z=0$ halos of mass
$M_{\rm{vir}}\sim10^{13}$ h$^{-1} M_\Sun$ formed at $z<1$ while in
halos of mass $M_{\rm{vir}}\sim10^{12}$ h$^{-1} M_\Sun$ $\sim80$\% of
the stars were formed over the same interval. This is generally
consistent with observed trends \citep[e.g.,][]{Heavens04, Noeske07}.
The implied star-formation rates from this comparison are sensitive to
the less massive end of the GSMF at $z=1$, which is less-well
constrained observationally, and should thus be treated with a
caution.  Note that the models converge at lower masses, and thus
these conclusions are insensitive to the way in which disrupted
subhalos are handled.

\subsection{Caveats \&  Assumptions}\label{s:robust}

\subsubsection{Definition of the BCG}

It has recently come to light that standard photometry in large galaxy
surveys systematically underestimates BCG luminosities for several
different reasons \citep{Lauer07, Bernardi07}.  This issue is
complicated by the somewhat arbitrary distinction between BCG and ICL,
as one needs a well defined notion of a BCG in order to claim that
standard techniques are ``missing'' BCG light.  The effect can be as
large as 1 magnitude, though in such cases it appears that the ICL is
included as part of the BCG.  Our results are robust against these
effects because our notion of a BCG is precisely that measured by the
data to which we compare \citep[i.e. by][]{Lin04b}, while the ICL is
simply light outside the optical radius (i.e. outside the region
counted as the BCG).  Unfortunately, this means that our results which
rely on the separation between the BCG and ICL are not directly
exportable to other observations of the BCG and ICL if such
observations separate these two observables in different ways.  As
mentioned previously, a more robust approach to this type of modeling
would be to present actual surface brightness \emph{profiles} which
could then be compared to any well-defined observational sample
\citep[see e.g.,][]{Rudick06}.

\begin{figure}
 \plotone{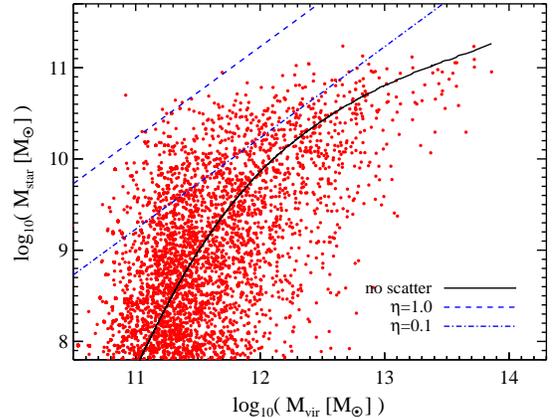}
\vspace{0.5cm}
\caption{Stellar mass as a function of virial mass at $z=1$.  The
  solid line indicates our standard one-to-one correspondence between
  halo and stellar mass, while the symbols represent a prescription
  for scatter between the two quantities (only $25$\% of the total
  number of objects are plotted for clarity).  The dashed and
  dot-dashed lines represent the stellar mass that a galaxy would have
  if it had a star-formation efficiency of $\eta=1.0$ and $0.1$,
  respectively (assuming that the baryon-to-dark matter ratio in each
  halo is the cosmic mean value of $0.17$). See the text for details.}
\vspace{0.5cm}
\label{f:scatter}
\end{figure}

\subsubsection{Scatter in the galaxy-halo connection}

We now explore the impact of scatter in the $M_{\rm{star}}-
M_{\rm{vir}}$ relation on our results (recall that the relation
between stellar mass and halo mass utilized in the previous section
was generated by assuming a one-to-one correspondence).  In order to
explore the maximal effect that scatter can have on our results, we
generate a rather extreme prescription of scatter.  The scatter is
included by multiplying each halo mass by a random number drawn from a
Gaussian with $\sigma=0.6$dex \citep[see][for a different prescription
of scatter]{Tasitsiomi04}.  Stellar masses are then matched to this
random number in the standard way\footnote{If $\zeta$ denotes the halo
  mass multiplied by a random number drawn from a Gaussian, then the
  stellar masses are assigned via $n_g(>M_{\rm{star,i}}) =
  n_h(>\zeta_{\rm{i}})$, c.f. Equation \ref{eqn:v2l}.}.  Figure
\ref{f:scatter} compares this model (solid symbols) to our standard
one-to-one correspondence (solid line).  This figure includes both
distinct halos and subhalos; for the latter we use the virial mass at
the epoch of accretion, as before.  In addition, we include lines that
indicate the amount of stellar mass a galaxy would have if its halo
contained the cosmic mean baryon-to-dark matter ratio, $f_b=0.17$, and
it converted a fraction $\eta$ of those baryons into stars ($\eta$ is
often called the star-formation efficiency).

Comparison between the scatter prescription and the lines of constant
star-formation efficiency indicate why this prescription is extreme
--- there are galaxies which have $\eta\sim1$ and indeed some rare
cases where the baryon fraction in the halo exceeds $f_b$.  At
$z\sim0$ the star-formation efficiency is almost certainly $\eta<0.25$
\citep[][and decreases for increasing halo mass]{Mandelbaum06}, so we
have effectively scattered galaxies, at a fixed stellar mass, to halo
masses that violate, or almost violate constraints on both $\eta$ and
$f_b$.  Note that this type of scatter (scattering down in halo mass
at a fixed stellar mass) will likely cause the most change in the
observables discussed in the previous sections because massive
galaxies in less massive halos will have less violent accretion
histories since $z=1$ compared to more massive halos.

However, even with this large amount of scatter, our results remain
qualitatively unchanged, though our conclusions are not as strong.
For example, the model predictions for the $L_{\rm{BCG}}-M_{\rm{vir}}$
relation are lower by $\sim0.1$dex, with a more pronounced tail toward
lower $L_{\rm{BCG}}$ at lower masses.  Evolution in the GSMF is also
less by about the same amount.  The BCG+ICL luminosity-mass relation
still rules out model \Bs, and the ICL fraction still favors model \Cs.

\subsubsection{Observational Uncertainties}

In $\S$\ref{s:evgsmf} we explored the sensitivity of the model
predictions to the adopted $z\sim1$ GSMF; we now do the same for the
BCG-cluster mass relation presented in $\S$\ref{s:BCG}.  We have
checked by eye the BCG-cluster mass relation produced from model \A
for the 200 realizations of the $z\sim1$ GSMF and compared them to the
observations of \citet{Lin04b}.  Model \A matches the observations
approximately $\sim13$\% of the time (note that this fraction is the
same fraction of realizations of model \A that match the observational
constraints on evolution in the GSMF, see $\S$\ref{s:evgsmf}).  In
these cases either $M_\ast$ or $\phi_\ast$ (or both) are $>1\sigma$
below the mean GSMF Schechter parameters.  Recall that these
uncertainties are rough combined systematic and statistical
uncertainties and are thus likely upper bounds.  Further, the GSMF we
adopt from \citet{Fontana06} is a lower bound with respect to other
observations at $z\sim1$ (see Figure \ref{f:obs}).  Nevertheless, any
future revision of the GSMF downward by $\gtrsim 1\sigma$ (systematic
+ statistical) would weaken the constraints on the models
significantly. Clearly, our results and conclusions could be sharpened
with more accurate measurements of the GSMF in the future.

\subsubsection{Effects of cosmology}

In our simulations of the $\Lambda$CDM cosmology, the adopted
normalization of the power spectrum, $\sigma_8=0.9$, was somewhat
higher than recent constraints from the 3-year {\sl WMAP}
data: $\sigma_8\approx 0.75-0.8$ \citep{Spergel06}. A lower value of
$\sigma_8$ would imply that the same observed galaxy number density
(and stellar mass) corresponds to smaller halo circular velocity and
virial mass. There would also be more evolution in the abundance of
massive halos between $z=1$ and $z=0$, as halo formation times would
be shifted to more recent time. This would imply that the amount of
evolution would be larger and our conclusions would be stronger in a
lower $\sigma_8$ universe. However, later formation times also imply
that there would be less time for accreting halos to merge and
contribute to the growth of the BCG stellar mass. The relative
importance of these competing effects will have to be quantified in a
future analysis similar to the one presented here with simulations of
lower $\sigma_8$.

\section{Discussion}\label{s:disc}

\subsection{Implications}

We have explored four models for the dissipationless evolution of
galaxies since $z=1$.  These models were constructed to match the
galaxy stellar mass function (GSMF) at $z=1$ and differ only in the
fate of satellite galaxies when the subhalo, within which the
satellite is embedded, is disrupted.  Upon confronting these models
with various observations we have found that only a model in which a
significant fraction of stars ($\gtrsim80$\%) from disrupted subhalos
are transfered to the ICL (referred to as model \C above) is
consistent with data.  The failure of the other models provides
significant insight into the dissipationless evolution of galaxies.

A model in which all the stars from disrupted subhalos are transfered
to the central galaxy (model \As) is strongly ruled out both by
observations of the evolution in the GSMF and observations of the
$z=0$ $L_{\rm{BCG}}-M_{\rm{vir}}$ relation.  Such a model would only
be viable if the observed GSMF at $z\sim1$ were significantly revised
downward from the current lowest reported measurements.  The failure
of such a model implies that, if stars from disrupted subhalos are
transfered to the central galaxy, then they cannot be put in the
central regions where BCG luminosities are measured.  Such a
conclusion is corroborated by simulations of dissipationless (``dry'')
galaxy-galaxy mergers, which find that the resulting galaxy generally
becomes more extended rather than substantially brighter at the center
\citep[e.g.][although this conclusion depends on the orbital
parameters of the accreted satellite]{Boylan-Kolchin07}.

We have also explored an extreme model where satellites never disrupt
even when their subhalos do (model \Bs).  This model fails
dramatically when compared to observations of the combined luminosity
of the BCG and ICL, under the assumption that the ICL is built-up
predominantly at $z<1$.  Although this assumption appears justified in
light of recent hydrodynamical simulations, one should note that our
conclusion as regards model \B relies on this assumption.  In our
models the massive subhalos correspond to massive satellites and it is
the fate of these massive satellites that most strongly affects the
comparison to observations of the combined luminosities of the BCG and
ICL. The failure of this model thus strongly suggests that the
disruption of subhalos in our high-resolution $N$-body simulations
corresponds to the disruption of satellite galaxies, at least for the
most massive subhalos. This need not have been the case; recent
semi-analytic models (SAMs) have decoupled the dynamical evolution of
subhalos from satellites when the subhalo disrupts \citep{Croton06,
  Wang06}, and hence these models produce a significant population of
satellites with no identifiable subhalo (the so-called ``orphan''
population).

The failure of model \B strongly suggests that in fact satellites
disrupt when their subhalo disrupts, at least for massive satellites,
and hence any model (including the SAMs mentioned above) which fails
to tie the fates of massive satellites to their subhalos will likely
fail to reproduce the observed combined luminosities of the BCG and
ICL.\footnote{However, the simulations used in the current generation
  of SAMs have coarser resolution than the simulations used herein.
  It is thus difficult to make a fair comparison between the different
  approaches to handling the satellite --- subhalo correspondence.}
Moreover, the failure of model \B provides additional justification for
previous modeling where a one-to-one relation between galaxies and
halos extracted from $N$-body simulations has been assumed
\citep[e.g.][]{Kravtsov04, Tasitsiomi04, Conroy06a}.

The data instead favor models where most, if not all, of the stars
from disrupted satellites are deposited into the ICL.  Model \C puts
all stars from disrupted satellites into the ICL while model \D puts
only half into the ICL and the rest onto the BCG.  Although
comparisons with various observations favor model \Cs, the
uncertainties and assumptions discussed in previous sections suggest
that reality may lie somewhere in between these two models.  The
strongest discriminant between these scenarios is the fraction of
BCG+ICL light contained in the ICL; as elsewhere, model \C most
faithfully reproduces these observations.

There is a growing consensus that massive red galaxies were more or
less in place by $z\sim1$ \citep[e.g.][]{Wake06, Cimatti06, Bundy06}.
At first glance it appears difficult to reconcile this fact with
$\Lambda$CDM simulations which show that massive dark matter halos
(the very halos in which these massive galaxies likely reside) grow by
factors of $\gtrsim 3$ since $z=1$.  The success of model \C resolves
this tension by ``hiding'' the accreted stars in the ICL.
Observations of the evolution of the ICL at $z<1$ will be needed to
substantiate this picture.

The success of model \C provides us with further insight into the
nature of the ICL.  Observationally the ICL appears to have colors
consistent with the BCG and thus contains primarily of old stars
formed at $z>1$ \citep{Krick06, Gonzalez00, Zibetti05}.  Is this
consistent with model \Cs, where the ICL is built up by mergers with
the BCG at $z<1$?  The answer to this question is most likely yes,
because the subhalos that are disrupting at $z<1$ were accreted onto
the host halo at $z\sim1$ (the average accretion epoch, weighted by
the fraction of stellar mass brought in by the subhalo is $0.93$).  In
other words, the galaxies that are contributing stars to the ICL at
late times were part of the main halo at $z\sim1$ and hence could
reasonably have had their star formation truncated by one or more
cluster-specific processes (e.g. ram pressure stripping or
harassment) by $z\sim1$.  Thus these galaxies would be adding
primarily old red stars to the ICL when they disrupt.

There are several other observables which can provide additional tests
of these models.  Specifically, the number of galaxies, $N(M)$, and
the total cluster luminosity, $L_{\rm{tot}}(M)$, both a function of
cluster virial mass, provide independent constraints compared to the
observations explored herein.  However, these two observables are much
more sensitive to star formation since $z=1$ (because they include
lower mass galaxies), which has been neglected in these models.  Hence
in the present work we have not not included a comparison to these
observables because such a comparison would require additional, less
constrained assumptions.

\subsection{Comparison to Related Work}

Semi-analytic models governing the formation and evolution of galaxies
have proven capable of reproducing both strong and mild evolution of
massive galaxies since $z=1$.  The models of
\citet[][KW06]{Kitzbichler07} and \citet{Bower06} produce relatively
mild evolution in the massive end of the GSMF, although KW06 appear to
over-predict the abundance of massive galaxies at $z=0$.  Meanwhile,
\citet{DeLucia07} and \citet{DeLucia06} use a semi-analytic model very
similar to that used in KW06 and found that massive galaxies roughly
double in stellar mass since $z=1$.  This doubling in stellar mass
does not strongly affect the evolution in the GSMF presented in KW06
because these authors assume $0.25$~dex uncertainty in the observed
stellar mass estimates at $z>0$.  This assumed uncertainty, which is
likely an upper bound, increases the model GSMF at the massive end for
$z>0$, and hence masks the stronger intrinsic evolution in the model.
None of these models attempt to model the ICL, and all are quite
sensitive to their treatment of the merging of satellite galaxies (the
various possible treatments are not explored in these models) as well
as an array of model parameters.  For these reasons it is difficult to
draw general conclusions from these models.

Our approach most closely parallels that of \citet{Monaco06} who used
a semi-analytic model to follow the evolution of the GSMF.  These
authors artificially turned off their star-formation prescription in
order to follow the dissipationless growth of galaxies since $z=1$,
similar to what we do here.  When a satellite merges with a central
galaxy, they transfer a fraction, $f_{\rm{scatter}}$, of the
satellites stars to the ICL.  They found that $f_{\rm{scatter}}\geq
0.3$ resulted in evolution in the GSMF in agreement with observations.

Several models presented in the present work are closely related to the
scheme employed in \citet{Monaco06}.  In particular, our models \As, \Ds,
and \C are similar to their model with $f_{\rm{scatter}}=0.0$, $0.5$
and $1.0$, respectively\footnote{Note that in their model $10$\% of
  the stars in the ICL came from the tidal stripping of satellite
  galaxies --- a process which we have ignored in the present work.}.
In $\S$\ref{s:evgsmf} we showed that models \D and \C were indeed in
agreement with the observed evolution of the GSMF while model \A was
not, similar to the conclusions of \citet{Monaco06}.  However, in
$\S$\ref{s:BCG} and $\S$\ref{s:icl} we showed that model \D
overproduced BCG luminosities at $z=0$ and underproduced the fraction
of combined BCG and ICL light contained in the ICL, while model \C
successfully reproduced both of these observations.  We hence expect
that the model presented in \citet{Monaco06} would reproduce these
latter two observations only if they used $f_{\rm{scatter}}\sim1$.
Comparing models to BCG luminosities and ICL fractions vs. cluster
virial mass provides unique constraints relative to evolution in the
GSMF because these two observables directly probe the properties of
the most massive systems, while the high mass end of the GSMF is
sensitive to Poisson uncertainty and cosmic variance.

Each of the models presented herein make predictions for the
disruption rate of satellite galaxies.  \citet{MWhite07} used the
redshift-dependent clustering of galaxies from $z\simeq0.9$ to
$z\simeq0.5$ to constrain the disruption rate of satellite galaxies.
Over this time interval, these authors found that at least $1-2$
satellites brighter than $\gtrsim1.6 L_\ast$ per massive halo were
disrupted.  When focusing on satellites comparable to those in White
et al., we find on average $1.1$ satellites within massive halos have
disrupted between $z=1$ and $z=0.5$ (for models other than model \Bs,
since in that model satellites never disrupt).  Agreement between
these results provides a satisfying cross-check for both approaches.
On the one hand, we follow directly the evolution of subhalos in
simulations and hence have useful information regarding their
disruption.  Conversely, White et al. rely primarily on evolution of
the observed clustering of galaxies to constrain the disruption rate,
and hence make no assumptions regarding the relation between subhalos
and satellites.  When combined, these results provide further weight
to the idea that, over the mass ranges explored herein, disrupted
subhalos correspond to disrupted satellite galaxies.

Most recently, \citet{Purcell07} have modeled the build-up of the ICL
by tracing the fate of satellites in dark matter merger trees
generated from the Extended Press-Schechter formalism combined with a
detailed prescription for satellite stripping \citet{Zentner05}.  This
dark matter treatment is coupled with a simple stellar mass-dark
matter halo correspondence at $z=0$ and a simple prescription for
star-formation.  The ICL in their model is built up from disrupted
satellites; with a reasonable threshold for subhalo disruption, these
authors found trends similar to those reported here (since their
merger trees are based on analytic prescriptions, they were able to
extend their analysis to halos at considerably lower mass then
explored herein).  In particular, they found that the fraction of
BCG+ICL light contained in the ICL in massive halos is considerable
and in agreement with the observations of \citet{Gonzalez05}.  The
agreement between their scheme and ours is encouraging given the
substantial differences in the detailed treatment.

\section{Summary}\label{s:conc}

In this paper we investigated models for the dissipationless buildup
of massive central galaxies and the intracluster light, in the context
of merging histories for dark matter halos in high resolution
simulations of the currently favored $\Lambda$CDM model.  We used a
simple model for associating galaxies with dark matter halos and
subhalos at $z=1$, using the observed galaxy stellar mass function and
the mass function for halos and subhalos measured in simulations.  The
dissipationless evolution of galaxies in this model was tracked with
the merging history of the dark matter halos extracted from
simulations.  We then confronted this model with data on the evolution
of the galaxy stellar mass function and with the amount and fraction
of cluster light that is in brightest cluster galaxies (BCGs) and in
the ICL at $z=0$, investigating where the predictions from variations
in the fate of stars in merging galaxies could be distinguished.

We found that our model accurately reproduces a variety of observed
properties at $z=0$ if disrupted subhalos deposit most of their stars
into the intracluster light (ICL).  Other scenarios, either those in
which most of the stars are deposited onto the central BCG or in which
stars from disrupted halos are left behind as satellite galaxies, are
strongly disfavored by the data.  Such a scenario suggests that, while
BCGs do not appear to evolve strongly at $z<1$, the ICL surrounding
such galaxies is growing substantially over this epoch.  This scenario
is corroborated by high-resolution dissipationless simulations of
galaxy-galaxy mergers \citep{Boylan-Kolchin07} which find that
disrupted satellites preferentially build-up the outer envelope of
massive galaxies.

Although ideally one should distinguish between light bound to
galaxies and light that is dynamically bound only to the main halo, it
is worth noting that our analysis does not explicitly distinguish
between the outer parts of bright or cD galaxies and the ICL.  For the
purposes of this work ``BCG'' refers to that part of the central
galaxy's light that is captured in standard survey photometry,
operationally defined for most of the comparisons herein as light
above a surface brightness cut of $\mu_i\approx 23$ mag arcsec$^{-2}$,
while ``ICL'' refers to all cluster light that is not contained in
galaxies using these measurements.  Further work both on the
theoretical and observation side is needed to refine this distinction,
and care should be taken when comparing various studies, as there is a
wide variation in choices made for these definitions.

We emphasize that models for the formation and evolution of galaxies
must be seriously confronted with observations of the ICL, in addition
to more conventional observations such as the GSMF and the two-point
correlation function.  The ICL contains a significant, if still
somewhat uncertain, amount of stellar mass, and models which ignore
this component will either place too much stellar mass in resolved
galaxies or will fail to produce enough stars globally.

The success of this simple model lends weight to earlier implications
from clustering statistics that the resolution of the current
generation of $N$-body simulations is sufficient to resolve the bulk
of subhalos that correspond to observed satellite galaxies in
clusters.  Such a confirmation unleashes an exciting array of
possibilities for understanding the connection between galaxies and
dark matter halos.

\acknowledgments 

We thank Yen-Ting Lin and Anthony Gonzalez for providing their data in
electronic format and for help in its interpretation, Brant Robertson
for pointing out the importance of stellar mass-loss, Jeremy Tinker
for providing the virial mass definition conversions, and Mike
Boylan-Kolchin for stimulating discussions regarding the ICL.  We
additionally thank Zheng Zheng, Frank van den Bosch, and Martin White
for helpful comments on an earlier draft.  CC thanks the city of
Montr{\'e}al for its unrelenting hospitality during the early stages
of this work.

The simulations were run on the Columbia machine at NASA Ames and on
the Seaborg machine at NERSC (Project PI: Joel Primack). We would like
to thank Anatoly Klypin for running these simulations and making them
available to us. We are also indebted to Brandon Allgood for providing
the merger trees. AVK is supported by the National Science Foundation
(NSF) under grants No.  AST-0239759 and AST-0507666, by NASA through
grant NAG5-13274, and by the Kavli Institute for Cosmological Physics
at the University of Chicago. This work made extensive use of the NASA
Astrophysics Data System and of the {\tt astro-ph} preprint archive at
{\tt arXiv.org}.

\end{document}